\documentclass[twocolumn,showpacs,preprintnumbers,amsmath,amssymb]{revtex4}
\usepackage{graphicx}
\usepackage{dcolumn}
\usepackage{bm}
\usepackage{dpscolor}
\usepackage{epsfig}

\begin{document}


\title{
Beam-helicity asymmetry measurements in the Virtual Compton
Scattering reaction in the $\gamma^* N\rightarrow \Delta$ transition
at $Q^2=0.20$ (GeV/c)$^2$}

\author{N.F.~Sparveris$^1$\footnote{current address: Massachusetts Institute of Technology}\footnote{corresponding author,
e-mail address: sparveri@jlab.org}, P.~Achenbach$^2$, C.~Ayerbe
Gayoso$^2$, D.~Baumann$^2$, J.~Bernauer$^2$, A.M.~Bernstein$^4$,
R.~B\"ohm$^2$, D.~Bosnar$^5$, T.~Botto$^4$, A.~Christopoulou$^1$,
D.~Dale$^6$\footnote{current address: Idaho State University,
Department of Physics, Pocatello, Idaho 83209, USA}, M.~Ding$^2$,
M.O.~Distler$^2$, L.~Doria$^2$, J.~Friedrich$^2$,
A.~Karabarbounis$^1$, M.~Makek$^5$, H.~Merkel$^2$, U.~M\"uller$^2$,
I.~Nakagawa$^3$, R.~Neuhausen$^2$, L.~Nungesser$^2$,
C.N.~Papanicolas$^1$, B.~Pasquini$^8$, A.~Piegsa$^2$,
J.~Pochodzalla$^2$, M.~Potokar$^7$, M.~Seimetz$^2$, S.~\v Sirca$^7$,
S.~Stave$^4$\footnote{current address: Triangle Universities Nuclear
Laboratory, Duke University, Durham, North Carolina 27708, USA},
S.~Stiliaris$^1$, Th.~Walcher$^2$ and M.~Weis$^2$}

\affiliation{$^1$Institute of Accelerating Systems and Applications
and Department of Physics, University of Athens, Athens, Greece}

\affiliation{$^2$Institut fur Kernphysik, Universit\"at Mainz,
Mainz, Germany}

\affiliation{$^3$Radiation Laboratory, RIKEN, 2-1 Hirosawa, Wako,
Saitama 351-0198, Japan}

\affiliation{$^4$Department of Physics, Laboratory for Nuclear
Science and Bates Linear Accelerator Center,
\\ Massachusetts Institute of Technology, Cambridge, Massachusetts 02139}

\affiliation{$^5$Department of Physics, University of Zagreb,
Croatia}

\affiliation{$^6$Department of Physics and Astronomy, University of
Kentucky, Lexington, Kentucky 40206 USA}

\affiliation{$^7$Institute Jo\v zef Stefan, University of Ljubljana,
Ljubljana, Slovenia}

\affiliation{$^8$Dipartimento di Fisica Nucleare e Teorica,
Universit\`a degli Studi di Pavia, and INFN, Sezione di Pavia, Pavia,
Italy}

\date{\today}

\begin{abstract}
We report on new beam-helicity asymmetry measurements $(A_h)$ of the
H$(\vec{e},e^\prime p)\gamma$ reaction in the $\Delta(1232)$
resonance at $Q^2=0.20$ (GeV/c)$^2$. The measurements were performed
at MAMI and were carried out simultaneously with the measurement of
the H$(\vec{e},e^\prime p)\pi^0$ reaction channel. It is the lowest
$Q^2$ for which the $A_h$ for the virtual Compton scattering (VCS)
reaction has been measured in the first resonance region. The
measured asymmetries have been compared and have been found to be
well described by dispersion-relation (DR) calculations for VCS. The
sensitivity of the data to the Generalized Polarizabilities (GPs) of
the proton and to the amplitudes associated with the nucleon
excitation to the $\Delta(1232)$ has been explored through the DR
formalism. The measured asymmetries have been found to exhibit a
much higher sensitivity to the GPs while it is suggested that future
measurements of higher statistical precision will offer stronger
constraints to theoretical calculations and will provide valuable
insight towards a more complete understanding of the GPs of the
proton and of the $\pi N$ amplitudes.

\end{abstract}


\maketitle

A fundamental question of nuclear physics is the description of the
internal structure of the nucleon. While the small distance
structure of the nucleon is adequately described by point-like
quarks and gluons, its description at larger distance can involve
both quark and baryon-meson degrees of freedom. The study of the
structure of the nucleon can be pursued by the exploration of a
number of complementary reactions and through the measurement of
various type of observables. Polarization observables has been
proven a very valuable tool towards this direction in the recent
years. Such observables have been explored at higher energies to
access the generalized parton distributions through the exclusive
deep virtual compton scattering reaction and for the study of the
transverse spin structure of the nucleon in semi-inclusive deep
inelastic scattering while at lower energies they can provide
significant sensitivity to various amplitude interferences through
the exploration of pion- and photon-electroproduction reactions.

One of the reaction mechanisms sensitive to the electromagnetic
structure of the nucleon is the Virtual Compton Scattering (VCS)
reaction. VCS off the proton refers to the reaction $\gamma^*p
\rightarrow e\gamma p$, which is accessed experimentally through
photon electroproduction $ep \rightarrow ep\gamma$, and is the
coherent sum of the Compton process and the Bethe-Heitler (BH)
process. The BH process refers to the photon emission by the
incoming or outgoing electron and it adds coherently to the VCS
amplitude. The Compton amplitude decomposes into a Born term,
characterized by a proton in the intermediate state, and a Non-Born
term containing all other intermediate states; the Born part is
given in terms of nucleon ground state properties and the non-Born
part contains all nucleon excitations and meson-loop contributions.
The BH and Born amplitudes are entirely calculable, with the proton
electromagnetic form factors as inputs.

The VCS has been proven in the recent years a valuable reaction,
complementary to elastic scattering and to pion-production, which
has provided new insights on the nucleon internal structure.
Previous experiments, both below and above the pion threshold, have
focused on the study of the generalized polarizabilities (GPs) of
the proton \cite{vcs1,vcs2,vcs3,jass} and the exploration of the
resonant amplitudes associated with the proton excitation to the
$\Delta(1232)$ resonance \cite{spavcs}, through cross section
measurements, as well as on the study of the imaginary part of the
VCS amplitude through beam helicity asymmetry $(A_h)$
measurements~\cite{vcs4}. In the present work we present new beam
helicity asymmetry $A_h$ measurements in the $\Delta(1232)$
resonance region and at four-momentum transfer of $Q^2 = 0.20$
(Gev/c)$^2$.

\begin{figure}[h]
\centerline{\psfig{figure=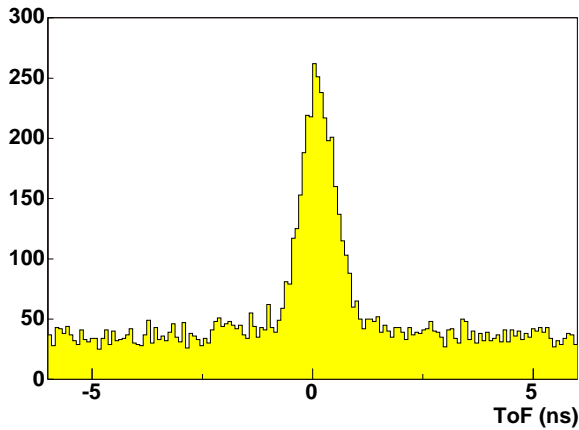,height=6.0 cm}}
\smallskip
\caption{The coincidence time spectrum for the H$(\vec{e},e^\prime
p)\gamma$ reaction.} \label{fig:tof}
\end{figure}

In order to measure the beam helicity asymmetry $A_h$ in the
$\vec{e} p \rightarrow ep\gamma$ process a longitudinally polarized
electron beam has to be utilized while the measurements have to be
performed above the pion threshold, where the VCS amplitude acquires
an imaginary part due to the coupling to the $\pi$N channel. Single
polarization observables, which are proportional to the imaginary
part of the VCS amplitude, become nonzero above the pion threshold.
The $A_h$ also requires measurements at out-of-plane values of
$\phi$, where $\phi$ is the reaction azimuthal angle with respect to
the momentum transfer direction and the scattering plane, due to the
sin$\phi$ type of dependence of the asymmetry. The beam helicity
asymmetry is defined as $A_h =
(\sigma^+-\sigma^-)/(\sigma^++\sigma^-)$ where $\sigma^+$ and
$\sigma^-$ designate the photon electroproduction cross-section with
beam-helicity state + and -, respectively. Since the BH and Born-VCS
contributions are purely real, the $A_h$ is due to the interference
of the imaginary part of the non-Born VCS amplitude with the real
BH+VCS amplitude. After development, one obtains in the nominator
the sum of a pure VCS contribution and a VCS-BH interference term
which has the effect to enhance the asymmetry. In the case of $A_h$
the interference of VCS with BH is desirable since it serves as an
amplifier of the measured asymmetry; the $A_h$ would be much smaller
in the absence of the BH process and the presence of VCS alone.

\begin{figure}[h]
\centerline{\psfig{figure=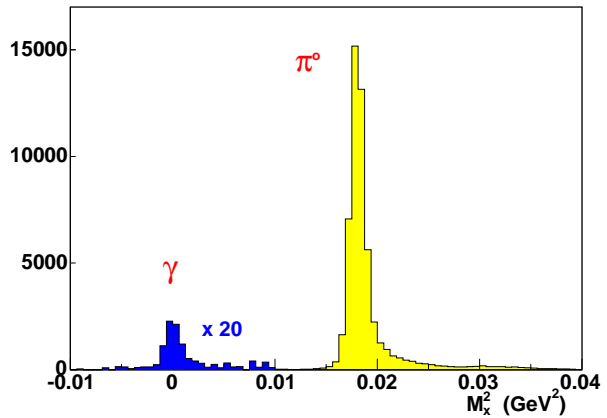,height=6.0cm}}
\smallskip
\caption{The derived missing mass spectrum, plotted as a function of
the square of missing mass shows the superb resolution achieved,
essential to isolating the small photon decay branch of the
$\Delta(1232)$ Resonance. Channels for $M_x^2 < 0.01 GeV^2$ have
been multiplied by a factor of twenty.} \label{fig:Mmiss}
\end{figure}

By measuring $A_h$ we can acquire information on the absorptive part
of the VCS amplitude and on the relative phase between the VCS
amplitude and the BH contribution \cite{kroll} aiming to a good
description of the VCS amplitude and to the understanding of its
$Q^2$ dependence. Through the $A_h$ measurements one can test the
theoretical models mostly towards their prediction of the imaginary
part of VCS. Towards this direction we exploit the
Dispersion-Relation (DR) calculation~\cite{dr1,dr2} for virtual
compton scattering. In this model the VCS non-Born contribution is
given in terms of dispersive integrals relating the real and
imaginary parts of the amplitude. The imaginary part is calculated,
through the unitarity relation, from the scattering amplitudes of
electro- and photo-production on the nucleon, taking into account
the dominant contribution from $\pi N$ intermediate states. The DR
model has two free parameters, $\Lambda_{\alpha}$ and
$\Lambda_{\beta},$ related to the dipole electric and magnetic GPs,
respectively, while the amplitudes for $\gamma^{(*)}p\rightarrow
\gamma \pi$ entering the unitarity relation are taken from the
MAID-2007~\cite{mai00} model. A measurement of $A_h$ offers a strong
cross-check of the dispersion formalism for VCS.

\begin{figure*}[t]
\centerline{\psfig{figure=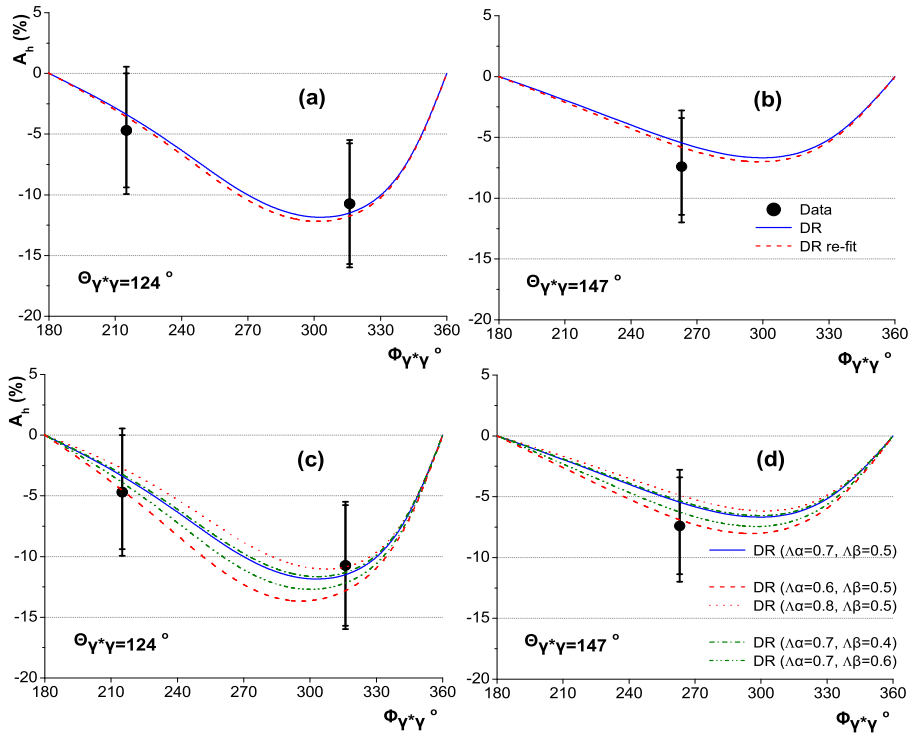,width=14.0cm,height=11.5 cm}}
\smallskip
\caption{The measured beam-helicity asymmetries $A_h$ as a function
of $\phi_{\gamma^*\gamma}$. The errors correspond to the statistical
and the total experimental uncertainties. The DR calculation is
presented along with the data. At the top panels the sensitivity to
the $\pi N$ amplitudes is explored (with the re-fit resulting from
\cite{spaplb}). At the bottom panels the sensitivity to the GPs is
investigated.} \label{fig:results}
\end{figure*}

The experiment was performed at the Mainz Microtron MAMI utilizing
an 855~MeV longitudinally polarized electron beam with an average
beam current of 25~$ \mu A$ employed on a 5~cm long liquid-hydrogen
target. The longitudinal beam polarization was $\approx 75\%$
throughout the experiment. The A1 magnetic spectrometers
\cite{spectr} were used for the proton and electron detection;
Electrons and protons were detected in coincidence with
spectrometers A and B respectively. The detector package in each
spectrometer includes a set of vertical drift chambers for particle
tracking and scintillator detectors for timing and particle
identification measurements while a Cerenkov counter was also used
for the electron identification. The H$(\vec{e},e^\prime p)\gamma$
reaction was performed at $Q^2=0.20$ (GeV/c)$^2$ and at
$W=1210$~MeV. The measurement of the H$(\vec{e},e^\prime p)\gamma$
reaction channel was performed simultaneously with the measurement
of the H$(\vec{e},e^\prime p)\pi^0$ channel \cite{spaplb} which was
the primary goal of the experiment. A detailed description of the
experimental arrangement and the parameters for all the measured
setups are reported in \cite{staveprc}. The data acquisition time
was optimized for the measurement of the $\pi^0$ channel and thus
high statistical accuracy was not achieved for the photon channel
measurements. The measurements were taken  for
$\theta_{\gamma^*\gamma}=124^\circ$ and $147^\circ$, with
$\theta_{\gamma^*\gamma}$ the polar angle in the c.m. frame between
the initial and final photons of the VCS process, and for a range of
azimuthal angles with respect to the electron scattering plane
$\phi_{\gamma^*\gamma}$ from $215^\circ$ to $316^\circ$; extracting
the $A_h$ at different azimuthal angles is useful towards the
understanding of the $\phi_{\gamma^*\gamma}$ dependence of $A_h$
which is not known analytically. The kinematics of the present work
compared to the ones of a previous similar measurement \cite{vcs4}
is lower in $Q^2$ ($0.20$~(GeV/c)$^2$ instead of $0.35$~(GeV/c)$^2$)
and complementary in the range of the measured polar angles
$\theta_{\gamma^*\gamma}$ since the present measurements were
performed at backward angles while the previous ones \cite{vcs4}
were taken at forward $\theta_{\gamma^*\gamma}$ (from 2.6$^{\circ}$
to 33.7$^{\circ}$).

The measured beam-helicity asymmetries $A_h$ are presented in
Fig.~\ref{fig:results}. Both the statistical and the total
experimental uncertainties are exhibited. The results are dominated
by the statistical uncertainties since the data acquisition time was
optimized for the measurement of the $\pi^0$ channel which is
characterized by a much higher count rate. The systematic
uncertainties are driven mainly by the uncertainty of the central
momenta and the spectrometer angles as well as from the uncertainty
introduced to the final results by the variation of the analysis
cuts. The central momenta and spectrometer angle uncertainties were
varied in the analysis procedure in order to quantify the
corresponding uncertainties. The deviations introduced to the
extracted values of $A_h$ resulting primarily from the variation of
the size of the kinematical phase space bins (polar and azimuthal
angles, invariant mass and four-momentum transfer), and of the
coincidence-time and missing-mass analysis cuts were quantified as
the corresponding analysis cuts systematic uncertainties. A cut on
the target length is also applied in the analysis to ensure the
elimination of any contribution resulting from background processes
coming from the target walls. Detector inefficiencies and dead-times
are canceled out in the asymmetry. Moller polarimeter measurements
provided the measurement of the longitudinal beam polarization with
a relative uncertainty better than 2\%; this uncertainty results to
a corresponding systematic uncertainty to the $A_h$. In order to
extract the final $A_h$ result a theoretical model has to be used to
project the finite phase space to point kinematics; for this part
the DR \cite{dr1,dr2} model was used. The projection introduces a
corresponding uncertainty due to the uncertainty of the description
of the theoretical model but this uncertainty is rather small, at
the order of $\approx0.1\%$ in absolute magnitude for the extracted
asymmetries. Radiative corrections \cite{rad1,rad2} have also a
negligible effect to the asymmetry. The overall systematic
uncertainties, for all of the experimental points, are approximately
50\% of the corresponding statistical uncertainty.

The results are compared to the predictions of the DR \cite{dr1,dr2}
calculation as exhibited in Fig.~\ref{fig:results}. The DR
calculation is found to describe successfully the data, although the
experimental errors are rather big. The $A_h$ has been extracted at
various $\phi_{\gamma^*\gamma}$ angles thus providing the
opportunity to explore its azimuthal angle dependence which is not
known analytically. The data support a
$sin\phi_{\gamma^*\gamma}$-like dependence as expected from the pure
VCS contribution in the numerator of the asymmetry. The sensitivity
of the DR calculation to both the $\pi N$ amplitudes and to the GPs
of the proton is also explored (top and bottom panels of
Fig.~\ref{fig:results}, respectively). The default DR calculation is
using the MAID $\pi N$ amplitudes; the calculation is also presented
with the values of the three $N \rightarrow \Delta$ resonant
amplitudes re-fitted utilizing the $\pi^0$ channel measurements
\cite{spaplb} at the same kinematics. As exhibited in
Fig.~\ref{fig:results}(a) and Fig.~\ref{fig:results}(b) the
sensitivity to the resonant amplitudes is quite small. A re-fit
including also the $S_{0+}$ amplitude, in addition to the three
resonant ones, was also performed; the sensitivity to the $S_{0+}$
amplitude is provided through the fifth-structure function
$\sigma_{LT}^{'}$ measurements of the $\pi^0$ channel \cite{spaplb}.
Although this re-fit differentiated significantly the value of
$S_{0+}$, the result of the corresponding DR calculation practically
coincided with the DR calculation utilizing only the resonant
amplitudes re-fit. The DR sensitivity to the GPs has also been
explored as exhibited in Fig.~\ref{fig:results} (panels (c) and
(d)). The DR calculation has been explored by applying a rather
small variation of $\pm~0.1$ to the $\Lambda_{\alpha}$ and
$\Lambda_{\beta}$ (the two parameters related to the dipole electric
and magnetic GPs respectively) with respect to the central
corresponding values of 0.7 and 0.5. It is evident that the $A_h$
offers a much higher sensitivity to the GPs compared to the $\pi N$
amplitudes. The highest sensitivity is exhibited to the electric
polarizability. Nevertheless, the uncertainty of the data is rather
big to allow us to significantly constrain these parameters. The
fact that our uncertainties are driven by statistical errors is
suggesting that future dedicated measurements of higher statistical
precision will be able to offer strong constraints to the electric
polarizabilitiy, at the present kinematics. It is quite interesting
that the interplay of the amplitudes to the GPs is quite different
in the present measurement compared to a previous measurement of
$A_h$ \cite{vcs4} at different kinematics; the previous results
\cite{vcs4}, which are at higher $Q^2$ kinematics and at forward
polar angles, exhibit higher sensitivity to the $\pi N$ amplitudes
compared to the GPs.

In conclusion we have performed measurements of the beam-helicity
asymmetry $A_h$ for virtual compton scattering in the $\Delta(1232)$
region accessing a new lowest $Q^2$. The experimental results, which
provide access to the imaginary part of the VCS, have tested the DR
formalism for VCS, where the imaginary part of the VCS amplitude is
connected through unitarity to the $\gamma^{*}\rightarrow \pi N$
amplitudes, and we have found that the calculation is in good
agreement with the data which show, although within rather large
uncertainties, a $sin\phi_{\gamma^*\gamma}$ dependence of the $A_h$.
The sensitivity to both the GPs and to the $\pi N$ amplitudes has
been explored, while also taking into account the simultaneously
measured pion electro-production channel at the same kinematics, and
the greatest sensitivity is exhibited to the generalized electric
polarizability $\alpha_{E}$. It is suggested that future $A_h$
measurements of higher statistical precision that will explore a
wide range of the [$Q^2$,$\theta_{\gamma^*\gamma}$] phase-space can
offer new valuable input towards a more complete description of the
GPs of the proton and of the $\pi N$ amplitudes.

We would like to thank the MAMI accelerator group for providing
the excellent beam quality required for these demanding
measurements. This work is supported at Mainz by the
Sonderforschungsbereich 443 of the Deutsche
Forschungsgemeinschaft (DFG) and by the program PYTHAGORAS
co-funded by the European Social Fund and National Resources
(EPEAEK II).

\end{document}